\documentclass[aps,prl,twocolumn,showpacs,groupedaddress]{revtex4}
\usepackage{graphicx}

\begin{document}

\title{Superconducting Fe-based compounds (A$_{1-x}$Sr$_x$)Fe$_2$As$_2$ with A = K and Cs with transition temperatures up to 37 K}
\author{Kalyan Sasmal$^1$, Bing Lv$^2$, Bernd Lorenz$^1$, Arnold Guloy$^2$, Feng Chen$^1$, Yu-Yi Xue$^1$, and Ching-Wu Chu$^{1,3,4}$}
\affiliation{$^{1}$TCSUH and Department of Physics, University of Houston, Houston, TX 77204, USA} \affiliation{$^{2}$TCSUH and Department of
Chemistry, University of Houston, Houston, TX 77204, USA} \affiliation{$^{3}$Lawrence Berkeley National Laboratory, 1 Cyclotron Road, Berkeley,
CA 94720, USA} \affiliation{$^{4}$Hong Kong University of Science and Technology, Hong Kong, China}
\date{\today }

\begin{abstract}
New high-T$_c$ Fe-based superconducting compounds, AFe$_2$As$_2$ with A = K, Cs, K/Sr and Cs/Sr, were synthesized. The T$_c$ of KFe$_2$As$_2$
and CsFe$_2$As$_2$ is 3.8 and 2.6 K, respectively, which rises with partial substitution of Sr for K and Cs and peaks at ~37 K for 50-60\% Sr
substitution, and the compounds enter a spin-density-wave state (SDW) with increasing electron number (Sr-content).  The compounds represent
p-type analogs of the n-doped rare-earth oxypnictide superconductors. Their electronic and structural behavior demonstrate the crucial role of
the (Fe$_2$As$_2$)-layers in the superconductivity of the Fe-based layered systems, and the special feature of having elemental A-layers
provides new avenues to superconductivity at higher T$_c$.
\end{abstract}

 \maketitle











Guided by the rule that high-temperature superconductivity usually occurs in strongly correlated electron layered systems as in the copper
oxides \cite{1}, Hosono's group started a few years ago to search for superconductivity in quaternary equiatomic rare-earth transition-metal
oxypnictides, ROTPn, where R = rare-earth, T =  transition-metal  and Pn = pnictogen. Indeed, superconductivity was found in ROTPn, where R =
La, T = Ni and Fe, Pn = P and As, with transition temperatures (T$_c$) up to 26 K in F-doped LaOFeAs \cite{2}. The observation generated immense
excitement due to the high T$_c$, and the significantly large amount of a magnetic component, Fe, which is considered antithetic to conventional
s-wave superconductivity. In the ensuing few weeks, after the initial report of T$_c$ = 26 K in La(O,F)FeAs, the T$_c$ was quickly raised to
41-52 K in other F-doped samples, R(O,F)FeAs, replacing La with other trivalent R with smaller ionic radii \cite{2,3,4,5,6,7}. This is
consistent with the reported positive pressure effect on the T$_c$ of Ce(O,F)FeAs \cite{8}. Thus, a new class of materials with a promising
potential for high T$_c$ that may rival the well-known cuprate high-temperature superconductors was born. Intensive studies followed to further
raise their T$_c$, and to unravel the underlying mechanism for superconductivity in R(O,F)FeAs. A subsequent high pressure study shows that the
pressure effect on the T$_c$ of Sm(O,F)FeAs depends on F-doping, i.e. positive when the sample is under-doped but negative when over-doped,
similar to the cuprates \cite{9}. The results suggest that the maximum T$_c$ of R(O,F)FeAs is around $\sim$55 K and higher T$_c$'s ($>$55 K) may
yet be discovered in compounds that are chemically different, but physically related to R(O,F)FeAs. We therefore examined the structurally
related layered system AFe$_2$As$_2$, with A = K, Cs, Sr, (K/Sr) or (Cs/Sr). We found KFe$_2$As$_2$ and CsFe$_2$As$_2$ exhibit superconducting
transitions at 3.8 K and 2.6 K, respectively. Furthermore, with Sr substitution, the T$_c$ of (K$_{1-x}$Sr$_x$)Fe$_2$As$_2$ and
(Cs$_{1-x}$Sr$_x$)Fe$_2$As$_2$ increases to a maximum T$_c$ of 36.5 K and 37.2 K, respectively, at x$\sim$0.5-0.6. A new family of Fe-based
layered compounds with a relatively high T$_c$ is thus discovered. Given that elemental K, Cs, (K/Sr) or (Cs/Sr)-layers separate the
(Fe$_2$As$_2$)-layers, this class of superconducting materials may provide new ways to raise T$_c$.

ROFeAs crystallize in the tetragonal ZrCuSiAs-type structure \cite{2,3} that consists of transition-metal pnictide (Fe$_2$As$_2$)-layers
sandwiched by rare-earth oxide (R$_2$O$_2$)-layers, as shown in Fig. 1a. Similar to the cuprate high temperature superconductors, the charge
carriers are supposed to flow within the (Fe$_2$As$_2$)-layers, and the (R$_2$O$_2$)-layers act as "modulation doping" layers while retaining
the structural integrity of the (Fe$_2$As$_2$)-layers. However, details of the layered structure of ROFeAs are different from the high T$_c$
cuprates: the formally divalent Fe is tetrahedrally coordinated to four As-atoms, whereas the divalent Cu in cuprates is coordinated to four
oxygens in a square planar manner. AFe$_2$As$_2$ (A=K and Cs) crystallize in the ThCr$_2$Si$_2$ structure type \cite{10,11}. It features
identical (Fe$_2$As$_2$)-layers as in ROFeAs, but separated by single elemental A-layers, as shown in Fig. 1b. In stacking the
(Fe$_2$As$_2$)-layers in AFe$_2$As$_2$, the layers are oriented such that the As-As distances between adjacent layers are closest. Nevertheless,
interlayer As-As distances in AFe$_2$As$_2$ are effectively nonbonding. In ROFeAs, adjacent (Fe$_2$As$_2$)-layers are stacked parallel, with
identical orientations, and the (Fe$_2$As$_2$)-layers are further isolated by more complex (La$_2$O$_2$)-slabs.

We have undertaken a systematic study of the (K$_{1-x}$Sr$_x$)Fe$_2$As$_2$ for x = 0, 0.1, 0.3, 0.5, 0.6, 0.7, 0.8, 0.9, and 1.0. In addition,
representative superconducting phases of (Cs$_{1-x}$Sr$_x$)Fe$_2$As$_2$ with x = 0.5 and 0.6, as well as CsFe$_2$As$_2$, were studied. All
ternary compounds were prepared by high-temperature solid state reactions of high purity K, Cs and Sr with FeAs. Phase-pure FeAs powder was
prepared from the reaction of pure elements in sealed quartz containers at 600-800 $^\circ$C. Polycrystalline samples of the title compounds
were prepared and handled under purified Ar atmosphere. Samples were prepared as follows: stoichiometric amounts of the starting materials were
mixed and pressed into pellets. The pellets were sealed in welded Nb tubes under Ar. The reaction charges were jacketed within sealed quartz
containers, and then heated for 20-24 hrs at 1000, 950, and 700 $^\circ$C for SrFe$_2$As$_2$, KFe$_2$As$_2$, and CsFe$_2$As$_2$, respectively.
In addition, SrFe$_2$As$_2$ was preheated at 1200 $^\circ$C for 1.5 hours, and CsFe$_2$As$_2$ was preheated at 550 $^\circ$C for 12 hours. For
the mixed-metal samples, (K,Sr)Fe$_2$As$_2$ and (Cs,Sr)Fe$_2$As$_2$, stoichiometric amounts of the ternary iron arsenides were thoroughly mixed,
pressed and then annealed within welded Nb containers (jacketed in quartz) at 900 $^\circ$C for 20-30 hours. The SrFe$_2$As$_2$ and the mixed
metal, (K-Sr) and (Cs-Sr), compounds are stable to air and moisture. However, KFe$_2$As$_2$ and CsFe$_2$As$_2$ are air- and moisture-sensitive.
The resulting polycrystalline samples were investigated by powder X-ray diffraction. XRD data of the end compounds (i.e. x = 0.0 and 1.0), shown
in Fig. 2, can be completely indexed to the tetragonal ThCr$_2$Si$_2$ structure. The refined tetragonal cell parameters of the isostructural
mixed-metal phases show a trend in cell volume that agrees with the atomic radii of the metals, i.e. cell volumes increase with increasing
alkali metal content. In addition, the c/a ratios changes significantly with Sr-incorporation in that the ratio decreases with increasing
Sr-content, while the a-parameter nearly remains unchanged. The contraction in the c/a ratio is most significant in the Cs-compounds. This
implies that the interlayer distance between the (Fe$_2$As$_2$)-layers and the relevant As-As distances decrease with Sr content. $\varrho$(T)
was measured by employing a standard 4-probe method using a Linear Research LR-700 ac bridge operated at 19 Hz and the magnetic field effect on
$\varrho$ was measured using a Quantum Design PPMS system for temperatures down to 1.8 K and magnetic fields up to 7 T. The temperature
dependence of the dc-magnetic susceptibility $\chi$(T) was measured using Quantum Design SQUID magnetometer at fields up to 5 T. The Seebeck
coefficient was measured using a very low frequency ac two-heater method \cite{16}.

The resistivities $\varrho$(T), as a function of temperature and magnetic field; magnetic susceptibilities $\chi$(T) as a function of
temperature; and Seebeck coefficients of the title compounds were measured. The $\varrho$(T)s of all samples investigated exhibit metallic
behavior. Fig. 3a shows that the $\varrho$ of SrFe$_2$As$_2$ decreases from room temperature and undergoes a rapid drop at $\sim$200 K,
indicative of the onset of a SDW state, similar to the isoelectronic BaFe$_2$As$_2$ \cite{17,18}. The observed noise near room temperature is
associated with the condensation of moisture in the samples. Fig. 3a also shows $\varrho$ of KFe$_2$As$_2$ decreasing with temperature, but with
a strong negative curvature, suggesting strong electron-electron correlation. The $\varrho$ finally drops to zero below $\sim$3.8 K, indicating
a transition to the superconducting state (Fig. 3a inset).

All samples, except SrFe$_2$As$_2$ and K$_{0.1}$Sr$_{0.9}$Fe$_2$As$_2$, display bulk superconductivity as evidenced by the drastic drop of
$\varrho$ to zero and a large Meissner effect at T$_c$. The $\chi$(T) for KFe$_2$As$_2$ and CsFe$_2$As$_2$ show superconducting transitions at
$\sim$3.8 K and $\sim$2.6 K as shown in Fig. 4a (inset), respectively. The $\chi$(T) of the samples with highest T$_c$,
(K$_{0.4}$Sr$_{0.6}$)Fe$_2$As$_2$ and (Cs$_{0.4}$Sr$_{0.6}$)Fe$_2$As$_2$, are also shown in Fig. 4a. As expected, the magnetic field is observed
to suppress the superconducting transitions of (K$_{0.4}$Sr$_{0.6}$)Fe$_2$As$_2$ and (Cs$_{0.4}$Sr$_{0.6}$)Fe$_2$As$_2$, as shown in Fig. 3b.
Using the Ginzburg-Landau formula on Figs. 3b and 3c, and defining T$_c$ as the temperature at which $\varrho$ drops by 50\%, a high H$_{c2}$(0)
of 140 T and 190 T can be deduced for (Cs$_{0.4}$Sr$_{0.6}$)Fe$_2$As$_2$ and (K$_{0.4}$Sr$_{0.6}$)Fe$_2$As$_2$, respectively. Even at the 90 \%
resistivity drop the extrapolated H$_{c2}$(0) is still high, 110 T for (Cs$_{0.4}$Sr$_{0.6}$)Fe$_2$As$_2$ and 153 T for
(K$_{0.4}$Sr$_{0.6}$)Fe$_2$As$_2$. It is remarkable that these values exceed the critical fields of the fluorine-doped LaOFeAs compound
\cite{19}. The superconducting and the SDW transitions, evident from the $\varrho$(T) and $\chi$(T), is further confirmed by the measured
Seebeck coefficients presented in Fig. 4b. The significant positive thermoelectric power of (K$_{0.4}$Sr$_{0.6}$)Fe$_2$As$_2$ indicates the
major carriers in this system are hole-like (p-type), in contrast to the electron-like large negative thermoelectric power in the
superconducting CeO$_{0.84}$F$_{0.16}$FeAs.

The results of $\varrho$(T) and $\chi$(T) for phases with varying Sr content can be summarized by a phase diagram of T$_c$ vs. Sr content (x),
constructed for (K$_{1-x}$Sr$_x$)Fe$_2$As$_2$, in Fig. 5. A similar phase diagram for (Cs$_{1-x}$Sr$_x$)Fe$_2$As$_2$ is also observed. The
diagram shows that the T$_c$'s of KFe$_2$As$_2$ and CsFe$_2$As$_2$ are enhanced continuously by Sr-doping and peak at $\sim$37 K for Sr
substitution levels of 50-60\%.  In view of the valence counts of [(K)$^{1+}$]$_{0.5}$(FeAs)$^{0.5-}$ in KFe$_2$As$_2$ and
[(Sr)$^{2+}$]$_{0.5}$(FeAs)$^{1-}$ compared to (RO)$^{1+}$(FeAs)$^{1-}$ in ROFeAs, KFe$_2$As$_2$ exhibits significant electron deficiency,
whereas SrFe$_2$As$_2$ is isoelectronic to ROFeAs. Partial substitution of K by Sr corresponds to electron counts approaching the electron count
of ROFeAs, and the T$_c$ crests at doping levels corresponding to (FeAs)$^{(0.75-0.8)-}$. In contrast, the formal electron count corresponding
to the superconducting phase, RO$_{1-x}$F$_x$FeAs (x = 0.15-0.20), is (FeAs)$^{(1.15-1.20)-}$. Therefore, we conclude that the superconductivity
in the (Fe$_2$As$_2$) layers almost symmetrically peaks at two different types of carrier densities: p-type in (K/Cs,Sr)Fe$_2$As$_2$, and n-type
in R(O,F)FeAs. These observations demonstrate that Cooper pairs in the Fe$_2$As$_2$ layers can be formed by both holes and electrons, similar to
the behavior of the high-T$_c$ cuprates. The evolution of superconducting state to a SDW state by electron-doping in AFe$_2$As$_2$ and the
induction of a superconducting state from the SDW state in ROFeAs effectively demonstrates the symmetry in the extended phase diagram (hole and
electron carriers) and the unique role of (Fe$_2$As$_2$)-layers for superconductivity at relatively high temperature. Our results and
conclusions prove the significant role of the (Fe$_2$As$_2$)-layers in superconductivity of the Fe-based layer superconductors, and as the
primary cause for non-conventional superconductivity in these compounds. The simple AFe$_2$As$_2$ family of compounds also provides a basis from
which T$_c$ may be raised. This may be achieved by constructing more complex homologues of the layered Fe-pnictides, similar to what has been
carried out in the high-T$_c$ cuprates.

\begin{acknowledgments}
This work is supported in part by the T.L.L. Temple Foundation, the J.J. and R. Moores Endowment, the State of Texas through TCSUH, the USAF
Office of Scientific Research, and the LBNL through USDOE. A.M.G. and B.L. acknowledge the support from the NSF (CHE-0616805) and the R.A. Welch
Foundation. We also thank Zhongjia Tang for help with crystallographic calculations.
\end{acknowledgments}

\bibliographystyle{phpf}

\newpage

\begin{figure}
\includegraphics[angle=0, width=3.5in]{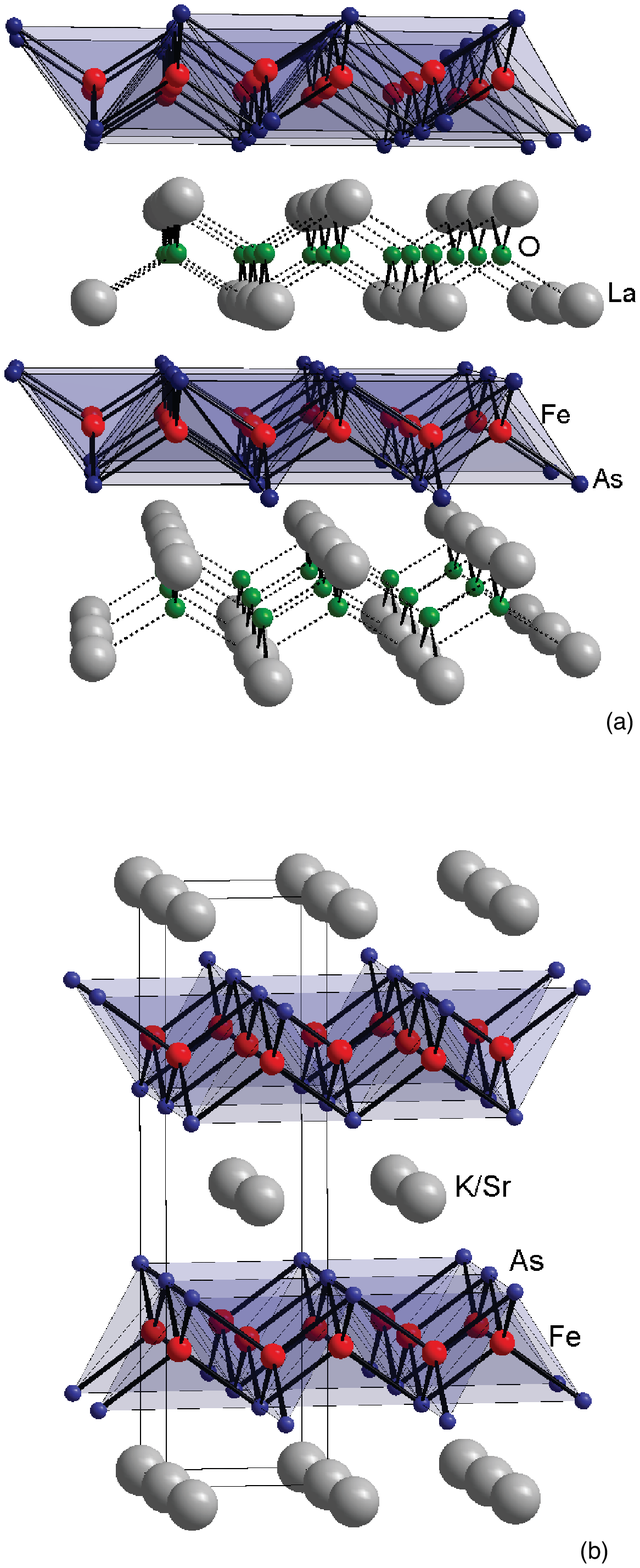}
\caption{(Color online) a) Crystal structure of LaOFeAs; b) crystal structure of (K/Sr)Fe$_2$As$_2$ and (Cs/Sr)Fe$_2$As$_2$.}
\end{figure}

\begin{figure}
\includegraphics[angle=0, width=3in]{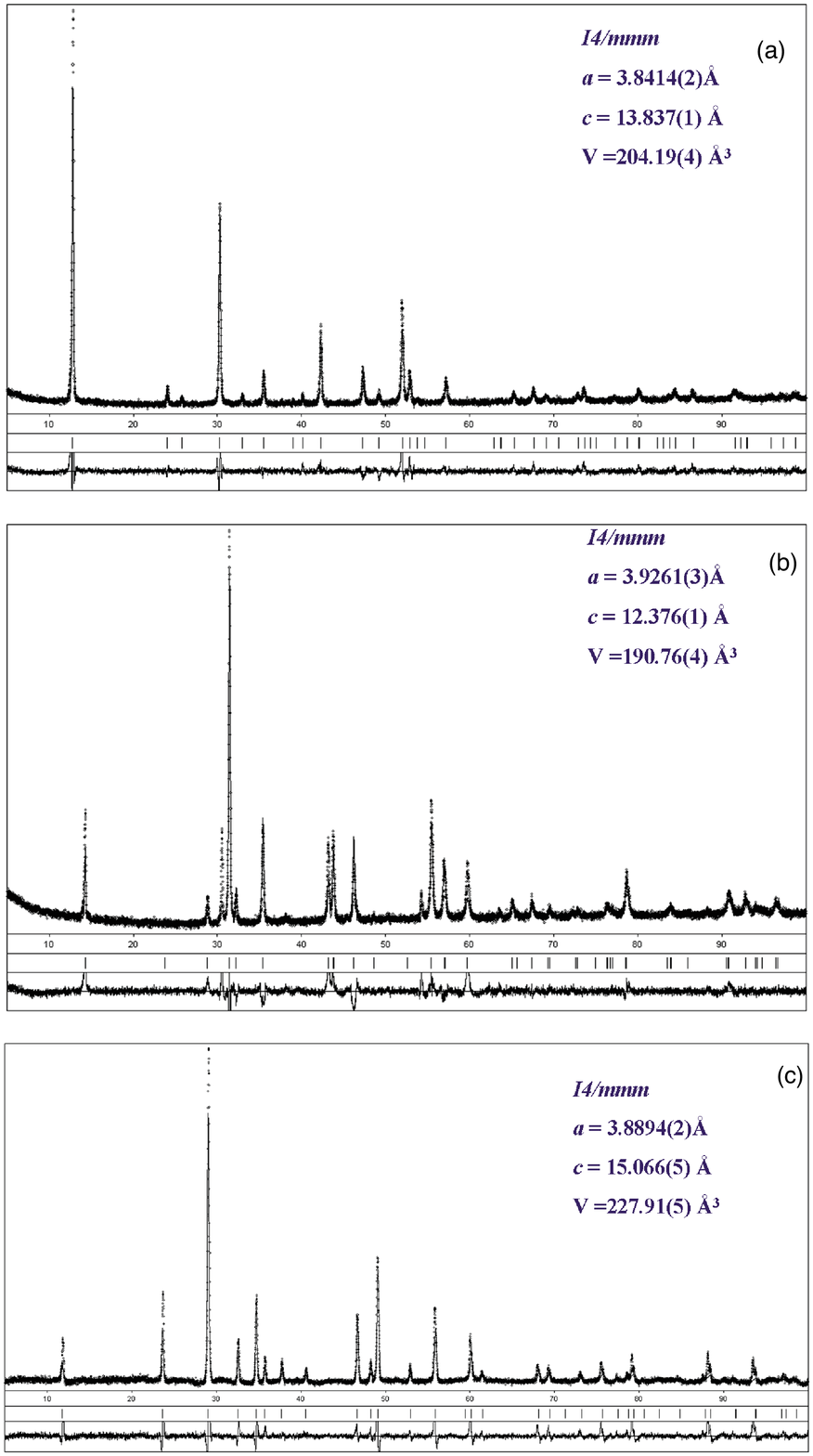}
\caption{a) Powder diffraction data/Rietveld refinement: a) KFe$_2$As$_2$, b) SrFe$_2$As$_2$, c) CsFe$_2$As$_2$.}
\end{figure}

\begin{figure}
\begin{center}
\includegraphics[angle=0, width=3in]{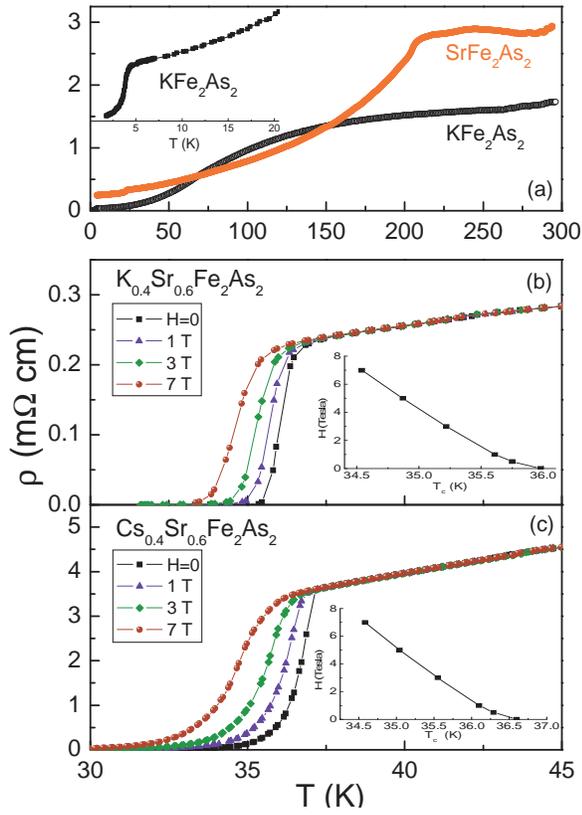}
\end{center}
\caption{(Color online) a) Resistivity of the two end members, SrFe$_2$As$_2$ and KFe$_2$As$_2$. The inset shows the superconducting transition
of KFe$_2$As$_2$ on an enlarged scale; b) Resistivity at different fields of K$_{0.4}$Sr$_{0.6}$Fe$_2$As$_2$; c) Resistivity at different fields
of Cs$_{0.4}$Sr$_{0.6}$Fe$_2$As$_2$. The insets in b) and c) show the T-dependence of the critical fields as determined from the midpoint of the
resistivity drop.}
\end{figure}

\begin{figure}
\begin{center}
\includegraphics[angle=0, width=3in]{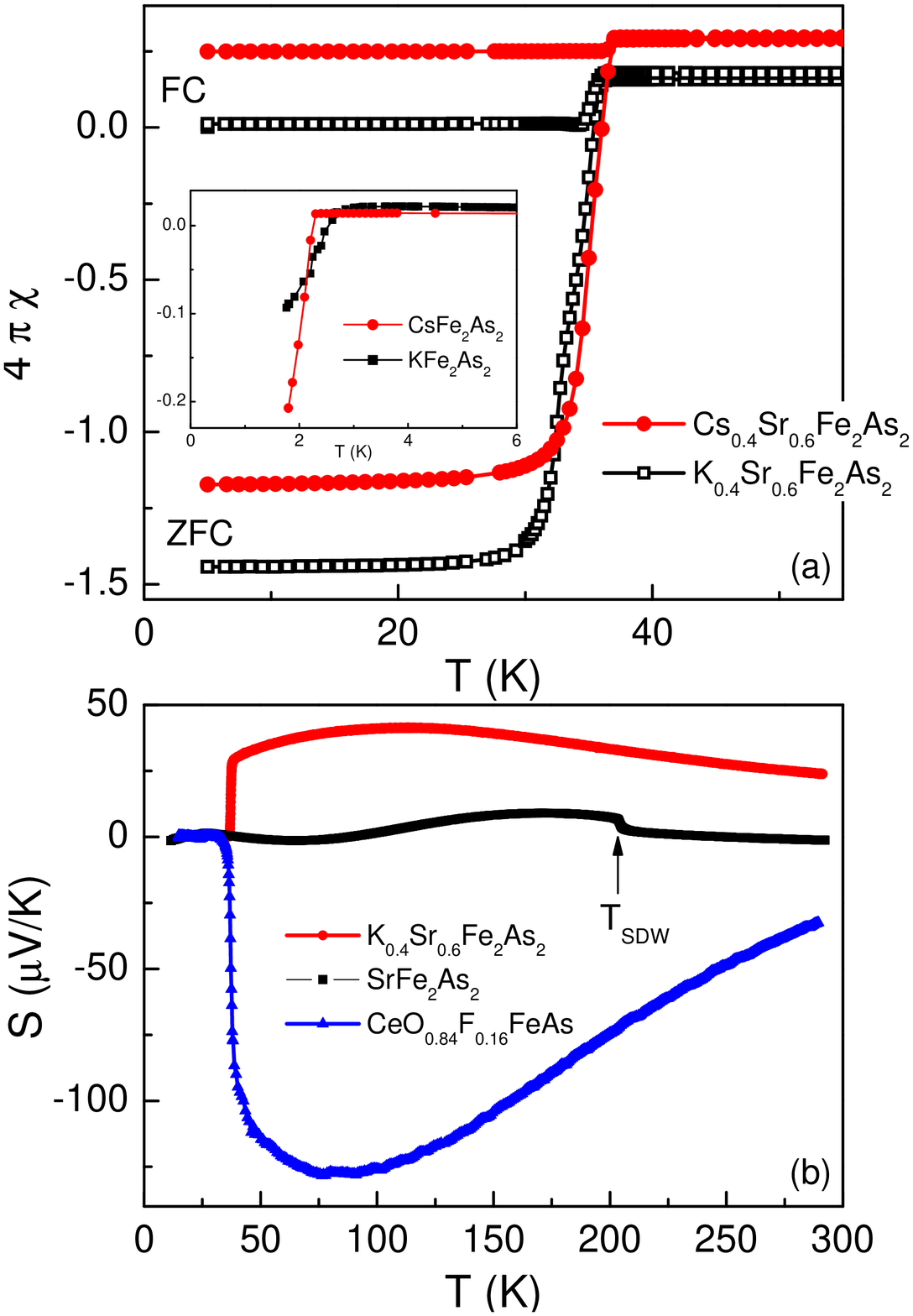}
\end{center}
\caption{(Color online) a) Magnetic susceptibilities of K$_{0.4}$Sr$_{0.6}$Fe$_2$As$_2$ and Cs$_{0.4}$Sr$_{0.6}$Fe$_2$As$_2$ measured at 10 Oe.
The inset shows the magnetic susceptibilities of KFe$_2$As$_2$ and CsFe$_2$As$_2$ near T$_c$; b) Seebeck coefficients of
K$_{0.4}$Sr$_{0.6}$Fe$_2$As$_2$ (top curve), SrFe$_2$As$_2$ (middle) and CeO$_{0.84}$F$_{0.16}$FeAs (bottom).}
\end{figure}

\begin{figure}
\begin{center}
\includegraphics[angle=-90, width=3in]{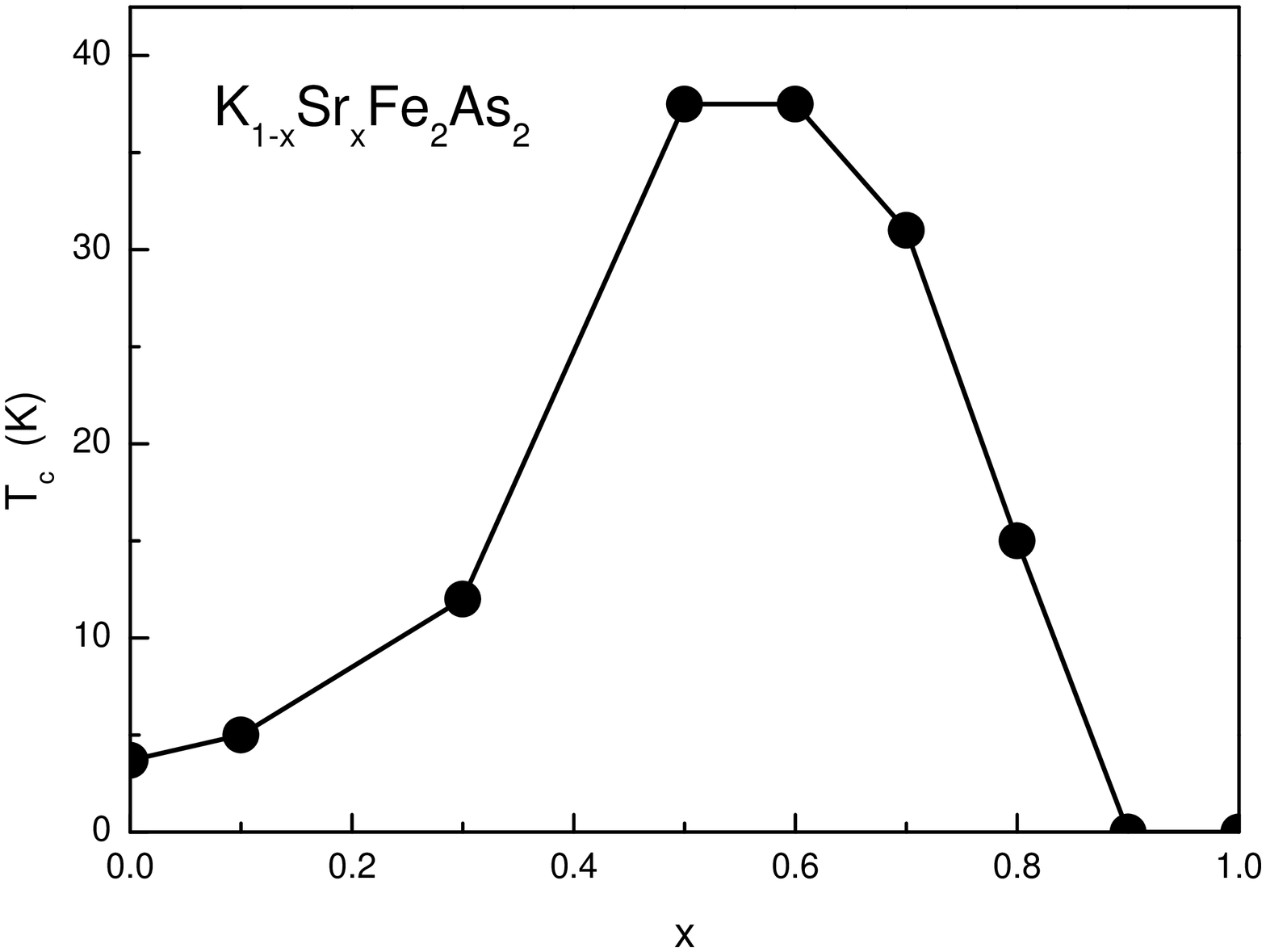}
\end{center}
\caption{Superconducting phase diagram of K$_{1-x}$Sr$_x$Fe$_2$As$_2$.}
\end{figure}

\end{document}